\documentclass[aps,prb,floats,twocolumn]{revtex4}
\usepackage{graphicx}
\usepackage{subfigure}
\usepackage{epsfig}
\usepackage{dcolumn}
\usepackage{bm}
\usepackage[ansinew]{inputenc}
\usepackage{amsmath}
\usepackage{amsthm}
\usepackage[T1]{fontenc}
\usepackage{amssymb}
\usepackage{amsfonts}
\usepackage[english]{babel}
\usepackage{enumitem}
\usepackage{color}

\begin{document}

\title{Switching of Magnetic Moments of Nanoparticles by Surface Acoustic Waves}

\author{J. Tejada,$^{1}$ E. M. Chudnovsky,$^{2,3}$ R. Zarzuela,$^{4}$ N. Statuto,$^{1}$ J. Calvo-de la Rosa,$^{1}$ P. V. Santos$^{5}$ and A. Hern\'{a}ndez-M\'{i}nguez$^{5}$}
\affiliation{$^{1}$Departament de F\'{i}sica de la Mat\`{e}ria Condensada, Facultat de F\'{i}sica, Universitat de Barcelona, Mart\'{i} i Franqu\`{e}s 1, 08028 Barcelona, Spain\\
$^{2}$Department of Physics and Astronomy, Lehman College of the City University of New York, 250 Bedford Park Boulevard West, Bronx, NY 10468-1589, USA\\
$^{3}$Graduate School of the City University of New York, 365 Fifth Ave, New York, NY 10016, USA\\
$^{4}$Department of Physics and Astronomy, University of California, Los Angeles, California 90095, USA\\
$^{5}$Paul-Drude-Institut f\"{u}r Festk\"{o}rpelektronik, Hausvogteiplatz 5-7, 10117 Berlin, Germany}

%\date{\today}

\begin{abstract}
We report evidence of the magnetization reversal in nanoparticles by surface acoustic waves (SAWs). The experimental system consists of isolated magnetite nanoparticles dispersed on a piezoelectric substrate. Magnetic relaxation from a saturated state becomes significantly enhanced in the presence of the SAW at a constant temperature of the substrate. The dependence of the relaxation on SAW power and frequency has been investigated. The effect is explained by the effective ac magnetic field generated by the SAW in the nanoparticles.
\end{abstract}


\maketitle

\section{Introduction}\label{introduction}
The recent years have witnessed a growing interest in the development of new schemes for the manipulation of magnetic moments. In this regard surface acoustic waves (SAWs) are well suited to trigger magnetic excitations in a local and selective manner.  For instance, magnetization dynamics due to SAW-induced magnetostriction in a ferromagnetic layer have been previously explored;\cite{Davis-APL2010,Kovalenko-PRL2013,Thevenard-PRB2013,Thevenard-PRB2014,Davis-JAP2015,Thevenard-PRB2016} the effect comes from the change in the magnetic anisotropy generated by the elastic strain induced by the SAW. Typical SAWs used in such experiments have frequencies in the ballpark of hundreds of MHz, and therefore the thickness of the ferromagnetic layer is large compared to the SAW wavelength $\lambda_{\textrm{SAW}}$. More recently, SAW-assisted magnetization switching from the single domain to the vortex state has been reported in mesoscopic Co elliptical disks lithographically patterned on a piezoelectric substrate.\cite{Sampath-Nano2016}

In this paper we investigate the possibility of switching magnetic moments of {\it nanoparticles} by SAWs. We find that an array of isolated magnetite nanoparticles dispersed on a substrate exhibits a significantly faster magnetic relaxation when subjected to a SAW. No temperature change in the substrate has been detected, thus ruling out the heating of the substrate by the SAW. On general grounds two mechanisms are expected to be responsible for these switching effects: i) magnetostriction on the magnetization due to the effect of the tensile stress on the magnetic anisotropy tensor and ii) the spin-rotation switching mechanism, which arises as a manifestation of the Einstein--de Haas effect at the nanoscale.\cite{chugarsch-PRB2005,garchu-PRB2015} The latter has been studied in the past in application to magnetic microcantilevers \cite{Kovalev-APL2003,Kovalev-PRL2005,Wallis-APL2006,jaachugar09prb,OKeeffe-PRB2013} and molecules sandwiched between conducting leads.\cite{JCG-EPL10} More recently it was proposed for the enhancement of the magnetization reversal by spin polarized currents\cite{Cai-PRB2014} and for achieving magnetization reversal by a pulse of the electric field in a torsional cantilever made of a multiferroic material.\cite{CJ-JAP2015}

We argue that field-induced magnetostriction can be discarded as the underlying mechanism for the observed effect: in contrast to ferromagnetic layers, the dispersed nanoparticles have a small contact area with the substrate---see e.g. Ref. \onlinecite{Boechler-PRL13}. The strain within the nanoparticle depends on the adhesion forces as well as on its mass and elastic properties, but the magnetoelastic coupling should be weak. In addition, the nanoparticles studied by us are very small compared to $\lambda_{\textrm{SAW}}$, and have, consequently, elastic resonance frequencies much higher than the SAW frequency. Finally, the quadratic dependence of the magnetization reversal on the SAW frequency---see Section \ref{theory}---makes the spin-rotation switching mechanism contribute dominantly within our experimental frequency range ---hundreds of MHz.\cite{Comm1} As a result, the nanoparticles will be displaced and rotated---rather than be just internally deformed by the SAW---, which leads to the generation of an effective ac magnetic field in the coordinate frame of the particle due to its local rotation.\cite{CJ-2016} Non-resonant absorption of the energy of this effective ac field by the particles is responsible for the enhanced magnetic relaxation. In retrospect, this mechanism may also be responsible for the spin dynamics generated by SAW in manganites \cite{Macia-PRB07} and molecular magnets.\cite{Macia-PRB08}

Theoretical works mentioned above deal with the mechanical oscillations at frequencies comparable to the frequency of the ferromagnetic resonance (FMR). It is well known---see, e.g., Refs. \onlinecite{Thirion-NatMat2003,CGC-PRB2013} and references therein---that the ac magnetic field of such frequency and a few Oe amplitude can reverse the magnetic moment via pumping of energy and angular momentum into the ferromagnetic system. Here we study SAW frequencies that are significantly lower than the FMR frequency of the nanoparticles. Nevertheless, as our experiments show and our theoretical analysis confirms, such SAWs are as effective in reversing magnetic moments of the nanoparticles.

The paper is structured as follows. Magnetic nanoparticles used in the experiment and the experimental setup are discussed in Section \ref{experiment}. Magnetic data are shown in Section \ref{magnetic}. Theoretical model and analysis of the data are presented in Sections \ref{theory} and \ref{analysis}, respectively. Section \ref{discussion} contains some final remarks and suggestions for further studies and possible applications.

\section{Experiment}\label{experiment}

%\begin{figure}[ht]
%\begin{center}
%\includegraphics[width=8.7cm,angle=0]{TEM.pdf}
%\caption{TEM of synthesized magnetite nanoparticles.}
%\label{TEM}
%\end{center}
%\end{figure}

\begin{figure}[ht]
\begin{center}
\includegraphics[width=8.7cm,angle=0]{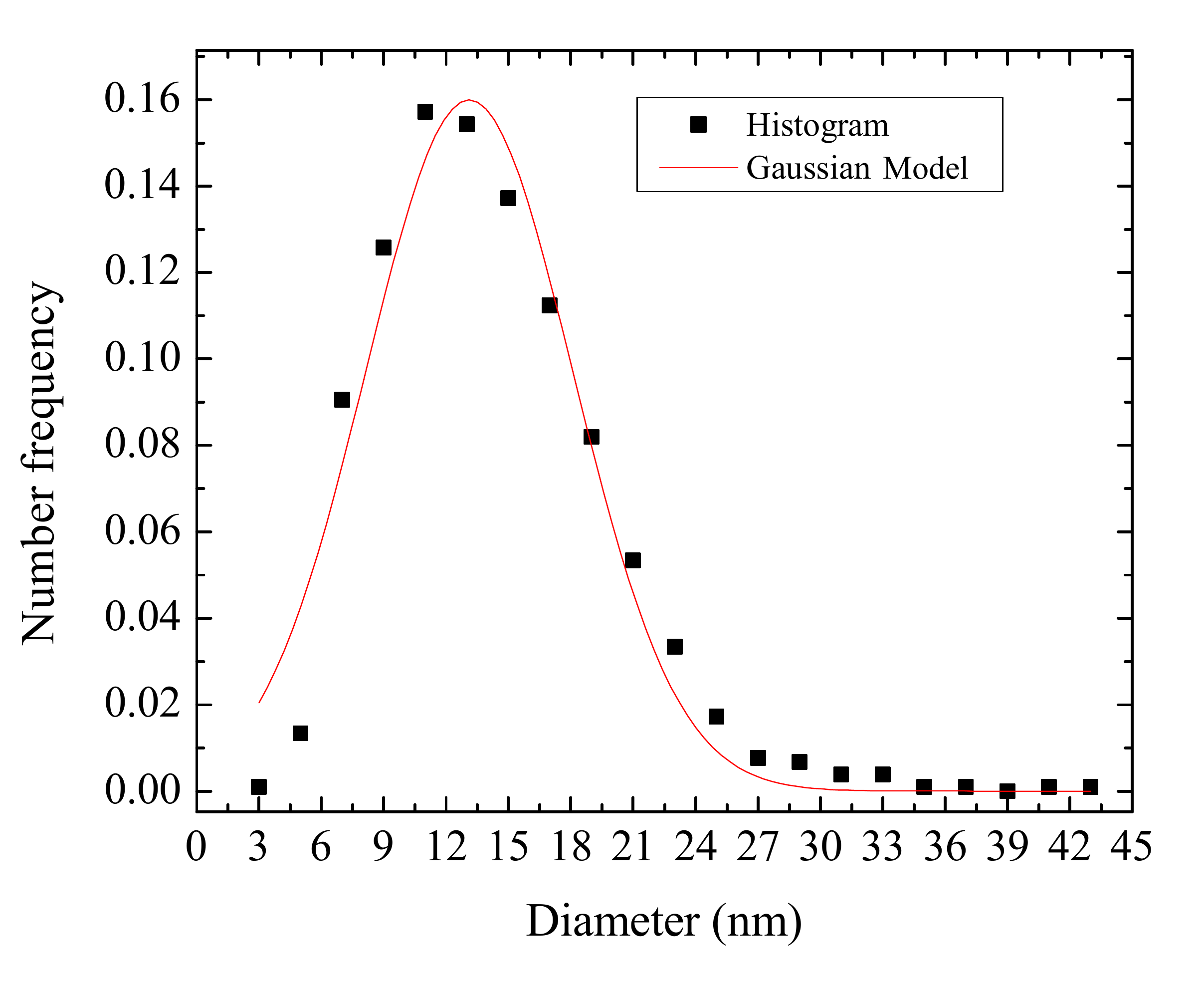}
\caption{TEM histogram of 1050 nanoparticles by number (black dots) and Gaussian fitting of the size distribution (red line). The corresponding values of the mean and standard deviation are $\mu=13$ nm and $\sigma=4.5$ nm, respectively.}
\label{Size}
\end{center}
\end{figure}

Magnetite nanoparticles (NPs) were synthesized via a procedure based on thermal decomposition of iron oleate at high temperatures in a high-boiling point organic solvent.\cite{Park2004}  This process started with the dissolution of a given amount of iron oleate in a given amount of solvent (long-chain hydrocarbon or amine). The resulting solution was heated up to a reaction temperature close to the boiling point of the solvent. After a given time at the reaction temperature, the solution was cooled down to room temperature. Transmission electron microscopy (TEM) images of the solution demonstrated the formation of (quasi-)spherical and triangular particles. The average size of the ribbon-shaped particles was calculated statistically from several TEM images by measuring the diameter of 1050 particles of the same sample, see Fig.\ \ref{Size}. 

\begin{figure}[ht]
\begin{center}
\includegraphics[width=9.0cm,angle=0]{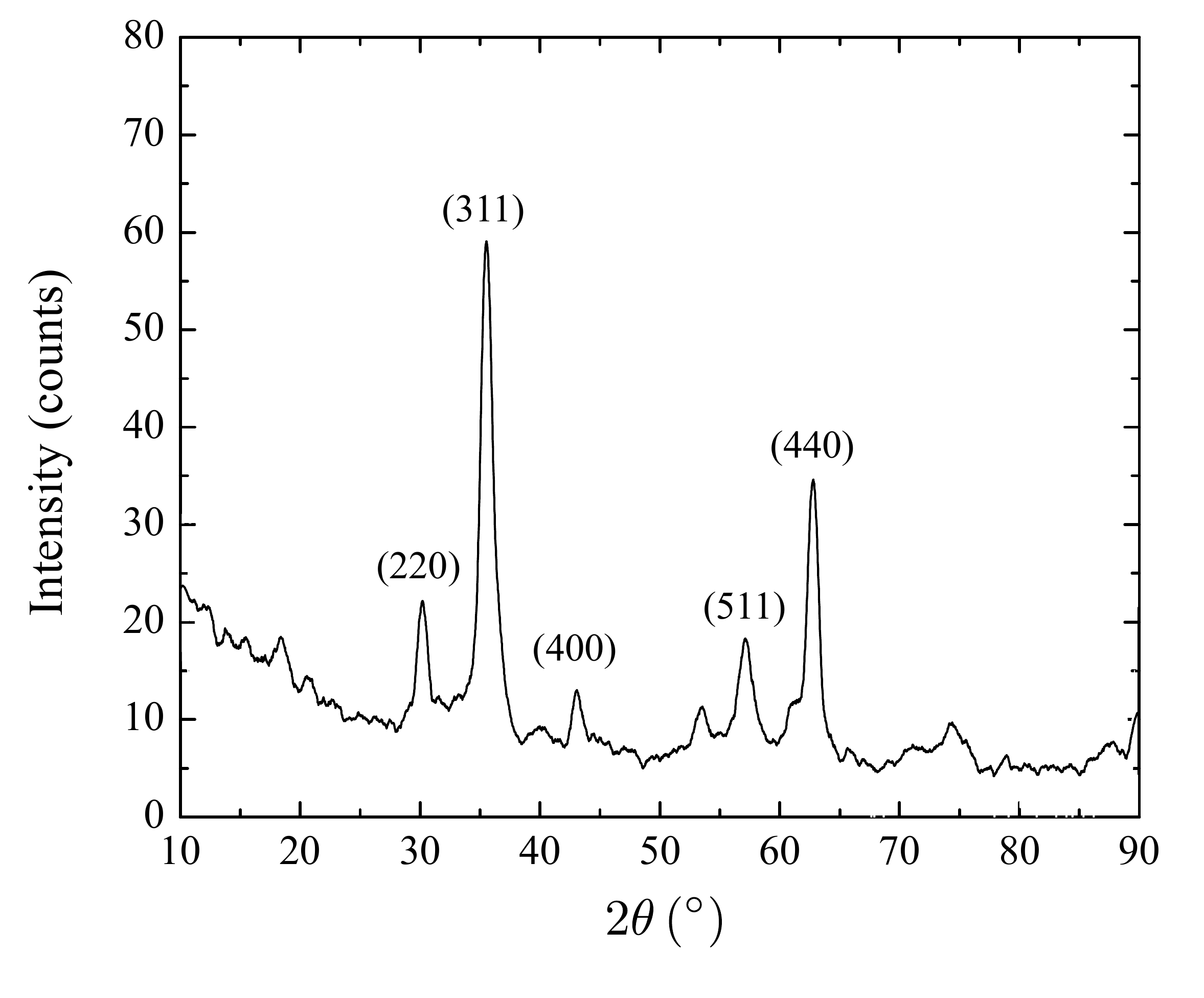}
\caption{X-ray difractogram of synthesized nanoparticles. Miller indices of magnetite are indicated for the main peaks.}
\label{X-Ray}
\end{center}
\end{figure}

The calculated mean size was 13 nm.  Magnetite powder was obtained by precipitation and separation from the solvent: $1$ ml of the colloidal dispersion of NPs was dissolved in $30$ ml of acetone, shaken, and centrifuged at $4000$ rpm for $20$ minutes. The supernatant was discarded and the precipitate containing the NPs was kept. The chemical composition, Fe$_{3}$O$_{4}$, was deduced from X-Ray powder diffraction (XRD) measurements, see Fig.\ \ref{X-Ray}. Narrow diffraction lines indicate high crystallinity of the particles. 

\begin{figure}[ht]
\begin{center}
\includegraphics[width=8.7cm,angle=0]{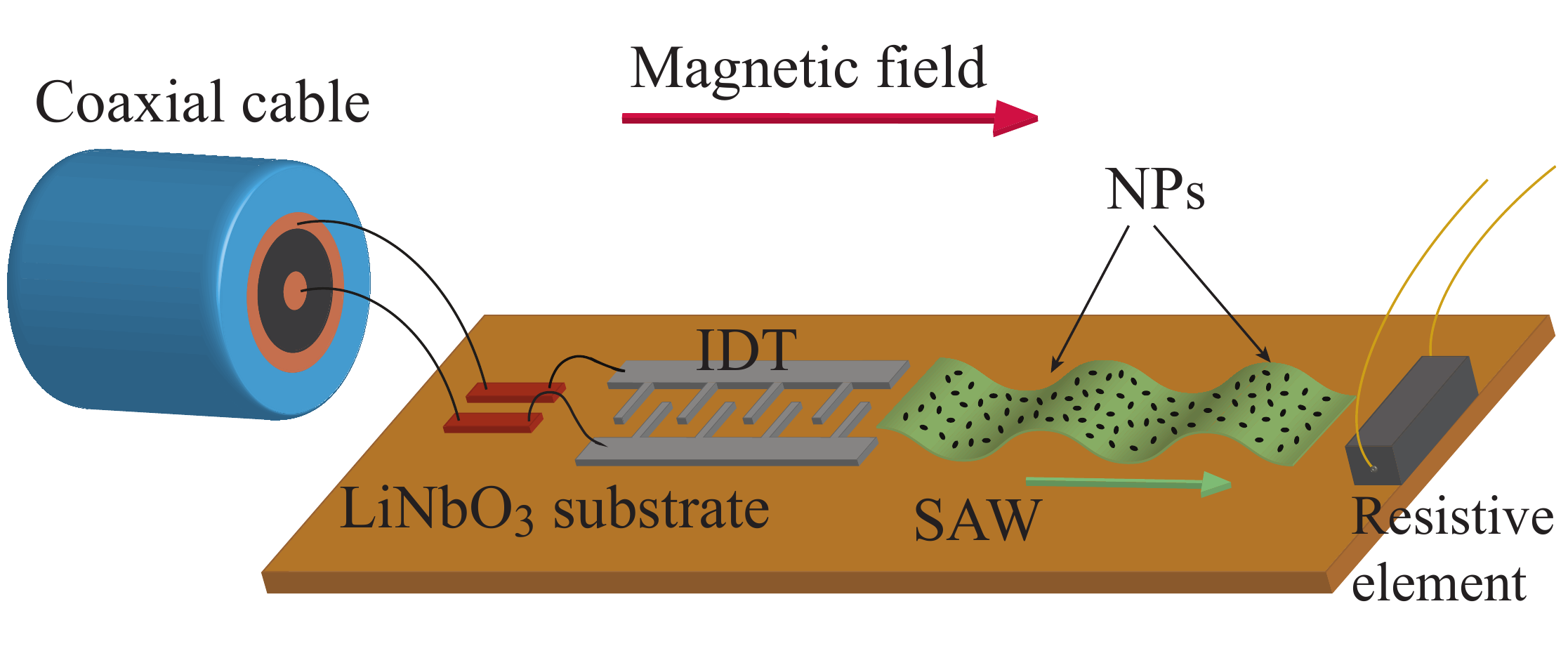}
\caption{Experimental setup.}
\label{setup}
\end{center}
\end{figure}

NPs have been deposited on a LiNbO$_{3}$, 128$^{\circ}$ Y-cut, substrate, see Fig.\ \ref{setup}. SAWs were generated by means of hybrid piezoelectric interdigital transducers (IDT) patterned on the same substrate. The spacing between the fingers was chosen to generate resonances with fundamental frequency $f_{0}=111$ MHz. Microwave excitation of the fundamental IDT resonance at $f_{0}=111$ MHz and of its harmonic at $3f_{0}=333$ MHz was achieved with the help of a commercial Agilent signal generator able to generate frequencies in the range from $250$ kHz to $4$ GHz. An Agilent network analyzer (PNA) was used to measure the reflection coefficient S$_{11}$ of the sample.
 
\section{Magnetic measurements}\label{magnetic}
\begin{figure}[ht]
\begin{center}
\includegraphics[width=8.7cm,angle=0]{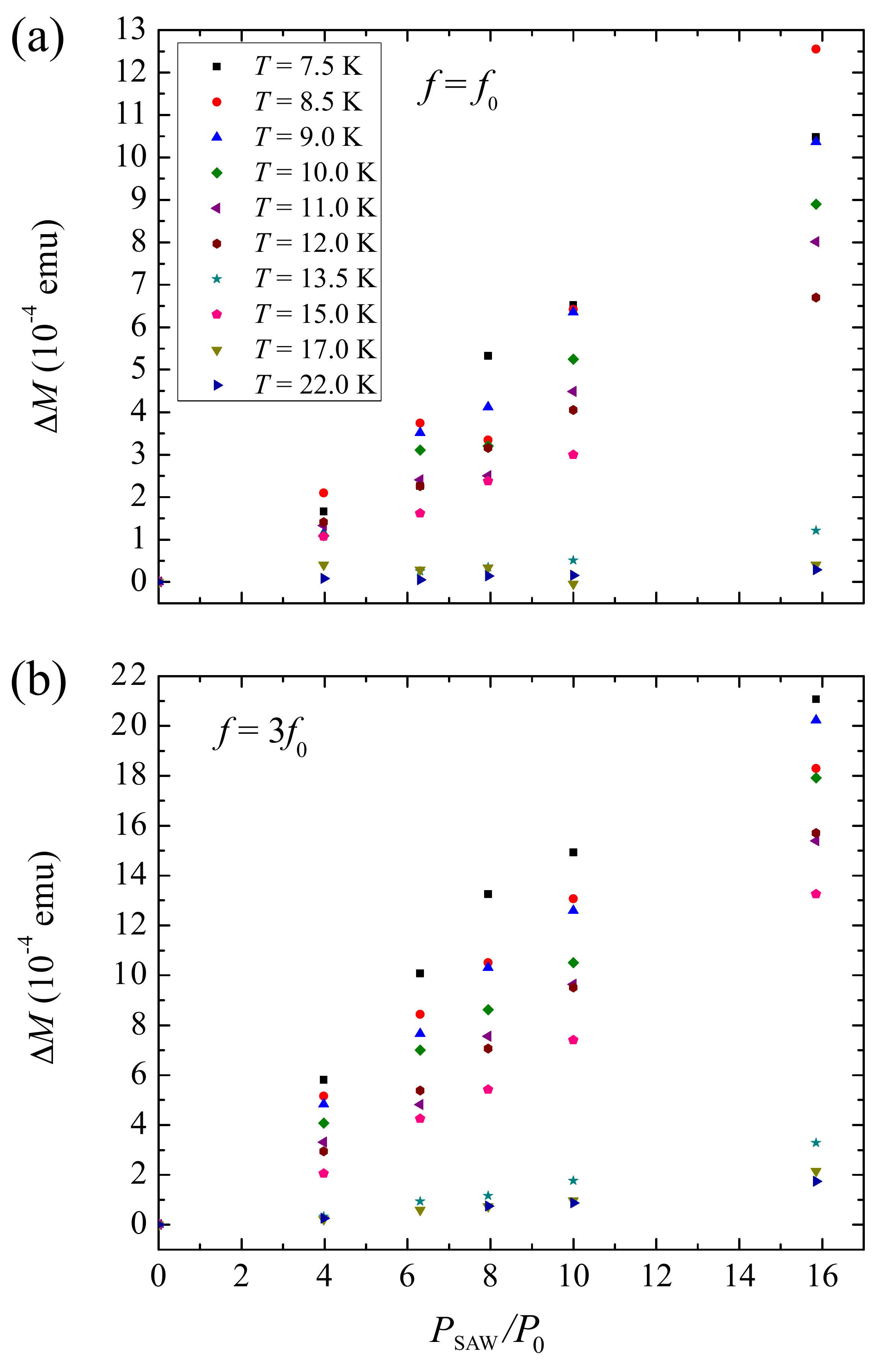}
\caption{Power dependence of the magnetization jump at the fundamental frequency of SAW, $f = f_0$, [panel (a)] and the third SAW harmonic, $f = 3f_0$ [panel (b)]. The power of SAW can be cast as $P_{\textrm{SAW}}=P_{\textrm{out}}\cdot\left[10^{\textrm{S}_{11}(\textrm{non-res})/10}-10^{\textrm{S}_{11}(\textrm{res})/10}\right]$, with $P_{\textrm{out}}=10^{L[\textrm{dB}]/10}$ mW being the power at the output of the generator. Here, $L$ is the decibel level of the output power, and S$_{11}$(res) and S$_{11}$(non-res) denote the values in decibels of the reflection coefficient S$_{11}$ at the resonant frequency and out of the resonance range, respectively. The size of the data points is bigger than the experimental error bars.}
\label{SAW}
\end{center}
\end{figure}

\begin{figure}[ht]
\begin{center}
\includegraphics[width=9.0cm,angle=0]{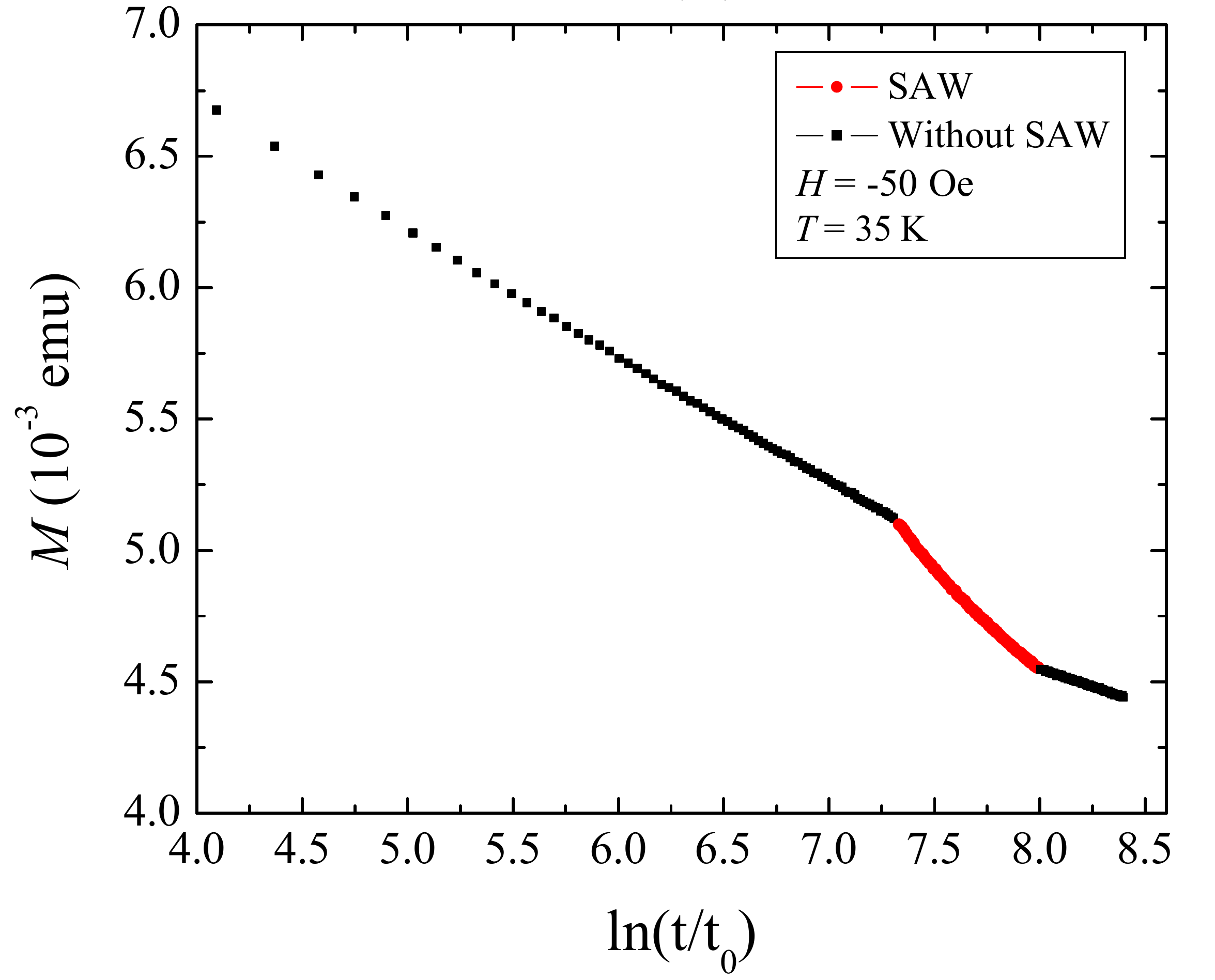}
\caption{ Isothermal ($T=20$ K) magnetic relaxation measurements from saturation ($H=5$ kOe) to $H=-50$ Oe in the presence (red dots) and the absence (black dots) of a SAW. The size of the data points is bigger than the experimental error bars.}
\label{MagSAW}
\end{center}
\end{figure}

Magnetic measurements were conducted inside a commercial rf-SQUID Quantum Design magnetometer by first saturating the sample of NPs in a positive magnetic field---oriented along the SAW propagation direction---of up to $5$ kOe and then reducing the magnetic field to zero, reversing it, and increasing the field in the opposite direction up to $-50$ Oe. Then magnetization of the sample was measured. This protocol was repeated at all temperatures ($2$ K -- $40$ K) in the absence and in the presence of a SAW, with a low-temperature stability better than 0.01 K and a magnetization sensitivity in the range of $10^{-8}$ emu.\cite{SQUID} Simultaneously, the frequency dependence of the reflection coefficient S$_{11}$ and of the magnetization of the NPs have been measured. Experiments with SAWs have been carried out at two available resonant frequencies, $f_{0}$ and $3f_{0}$---the second harmonic was strongly suppressed. Fig. \ref{SAW} shows the dependence of the corresponding magnetization jumps on the power of SAWs within the temperature range $7.5$ K -- $22$ K at these resonant frequencies. An analogous experimental protocol was followed along the ascending branch of the hysteresis cycle. Symmetric values of the magnetization were obtained at all temperatures in the absence and in the presence of a SAW. Field cooled (FC) and zero-field cooled (ZFC) magnetization curves were also measured in the interval $9$ K -- $300$ K at different applied magnetic fields along with isothermal magnetic relaxations along the descending branch from the saturation state, see Fig. \ref{MagSAW}. The effect of the SAWs on these measurements was to decrease the magnetization along the ZFC curve and to accelerate the relaxation dynamics.

The temperature of the system was measured with the SQUID thermometer. The temperature of the substrate was also independently measured by a resistive element attached to the substrate. Within their sensitivity limits, no heating of the substrate was observed when the SAW was applied. The reading of both thermometers coincided and remained constant in all individual measurements.

\section{Effect of SAW on magnetic relaxation of NPs}\label{theory}
The magnetic moment of a superparamagnetic particle of volume $V$ and magnetic anisotropy energy density $K$ switches due to thermal fluctuations at the rate $1/{\tau} =  (1/\tau_0) \exp\left[-{KV}/(k_BT)\right]$, where $\tau_0$ is the attempt time that is typically in the nanosecond range. The imaginary part of the ac susceptibility associated with these superparamagnetic transitions is given by 
\begin{equation}\label{chi}
\chi'' = \chi_0\frac{\omega\tau}{1 + (\omega\tau)^2}, 
\end{equation}
where  $\chi_0 = {M_{s}^2V}/{3k_B T}$ is the static equilibrium magnetic susceptibility, with $M_{s}$ being the saturation magnetization. Note that $\chi''$ peaks at $\omega = 2\pi f = 1/\tau$.

\begin{figure}[ht]
\begin{center}
\includegraphics[width=8.7cm,angle=0]{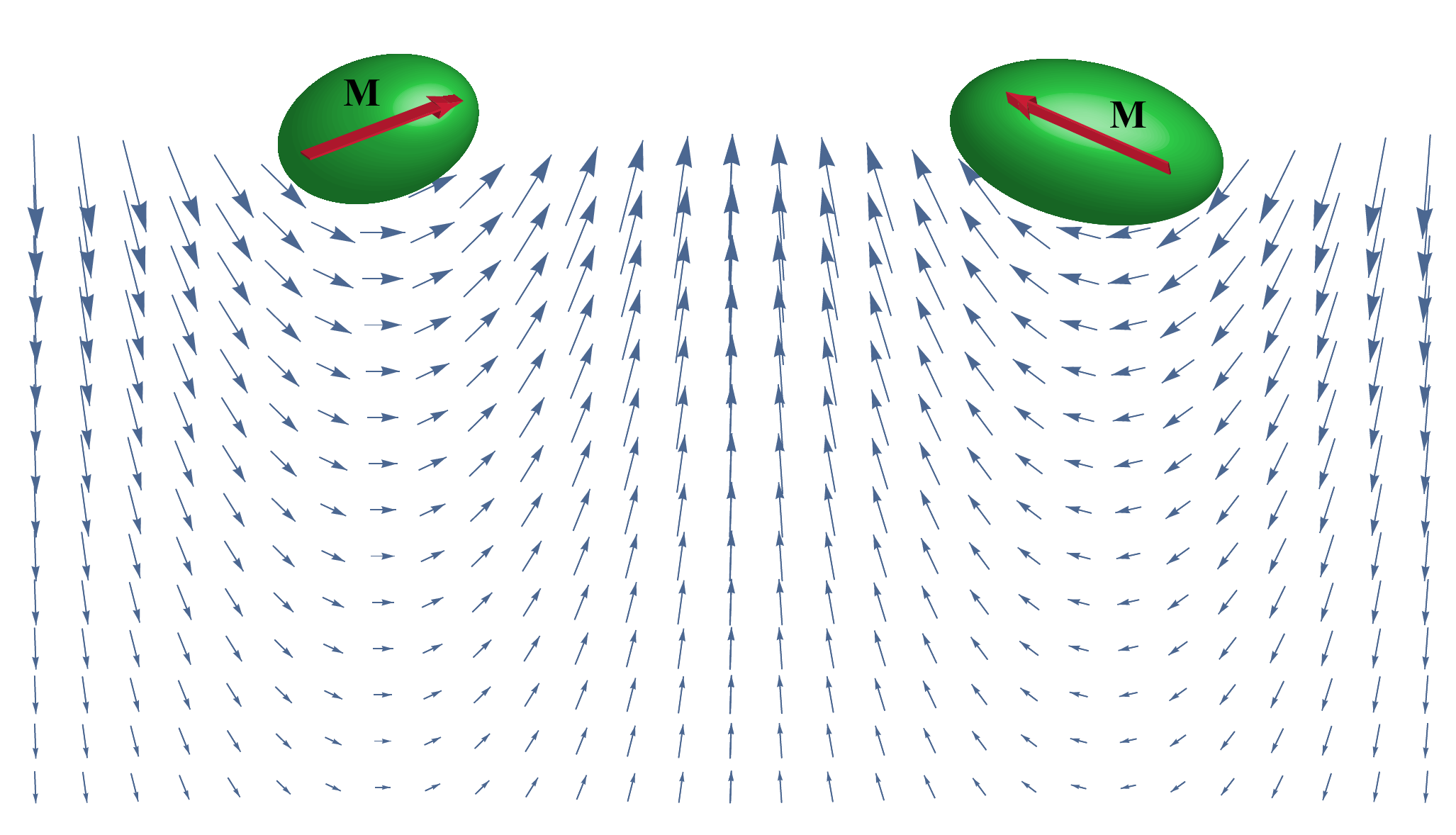}
\caption{Magnetic nanoparticles undergoing torsional oscillations due to surface acoustic waves. Blue arrows show the phonon displacement field generated by SAW in the substrate.}
\label{particle}
\end{center}
\end{figure}
For a nanomagnet rigidly adhered to a solid surface and subjected to SAWs the switching of the magnetization occurs in the coordinate frame coupled to the anisotropy axes of the magnet. SAWs generate rotations of the magnet---see Fig. \ref{particle}---and, thus, an effective ac field ${\bf h} = {\bm \Omega}/\gamma$ in the frame of the magnet. Here $\gamma$ denotes the gyromagnetic ratio, ${\bm \Omega} = \frac{1}{2} {\bm \nabla} \times \dot{\bf u}$ is the angular velocity, and ${\bf u}$ is the displacement field.  Only transverse displacements satisfying the condition ${\bm \nabla} \cdot {\bf u} = 0$ contribute to the mechanical rotation of the magnet. Transverse displacements from the source sending the SAW along the $X$-axis are given by \cite{LL}
\begin{equation}\label{u}
u_x = \kappa_t a \cos(kx -\omega t) e^{\kappa_t z}, \hspace{0.2cm} u_z = ka\sin(kx -\omega t) e^{\kappa_t z}
\end{equation}
where $a$ is related to the amplitude of the SAW and $\omega = c_t k \xi$ is the SAW frequency, with $\xi$ being a constant of order unity, $c_t$ being the speed of the bulk shear waves, and $\kappa_t =\sqrt{k^2 - {\omega^2}/{c_t^2}}$. In Eqs. \eqref{u} we have assumed that the substrate occupies the region $z\leq0$.

The angle of the local twist of the solid due to the SAW is ${\bm \phi} = \frac{1}{2}{\bm \nabla} \times {\bf u}$.\cite{LL} At the origin of the coordinate frame Eq.\ (\ref{u}) gives $\phi_x  = 0, \phi_y  =  -k^2a\xi^2\cos(\omega t), \phi_z  = 0$. The angular velocity is ${\bm \Omega} = \dot{\bm \phi}$, so that the effective ac magnetic field, ${\bf h} = {\bm \Omega}/\gamma$, felt by the nanomagnet at the origin is given by $h_x  = 0, h_y = h_0\sin(\omega t), h_z = 0$, with $h _0 = \omega k^2a\xi^2/\gamma$. We now notice that $ka = u_0$, where $u_0$ denotes the amplitude of the vertical displacement of the surface due to the SAW, and therefore $h_0 = \gamma^{-1}\omega k u_0 \xi^2 = {\xi} \omega^2 u_0/(\gamma c_t)$. In the actual experimental conditions, the amplitude $u_0$ lies in the ballpark of a few nanometers,\cite{Comm2} which is $10^{-4}$ of the wavelength $\lambda = c_t/f = 10\;\mu$m. Given the values $\xi \sim 1$, $\gamma = 1.76\times 10^7$ s$^{-1}$Oe$^{-1}$, $c_t=1.1\cdot10^5$ cm/s, $f =\omega/(2\pi)=111$ MHz and $u_{0}=4$ nm one obtains $h_0 \sim 0.1$ Oe for the effective magnetic field ---this value increases up to $0.3$ Oe for the third harmonic case. Since an ac magnetic field of this amplitude can strongly affect the magnetic relaxation,\cite{Comm3} this illustrates the importance of the spin-rotation mechanism.

The power of the effective ac magnetic field of amplitude $h_0$ and frequency $\omega$ absorbed per particle of volume $V$ at a temperature $T$ is given by \cite{Lectures} 
\begin{equation}\label{P}
P(V,T,f) = \pi z_p f h_0^2 \chi''(V,T,f) \propto  f^2 \chi''(V,T,f) P_{\textrm{SAW}}, 
\end{equation}
where $z_p \sim \lambda \propto 1/f$ is the penetration depth of the SAW into the layer of deposited NPs of thickness exceeding $z_p$, which was the case in the experiment. Here we have used the fact that the power of the SAW, $P_{\textrm{SAW}}$, scales as $ f^2 u_0^2$. As one turns the power on, the enhanced superparamagnetic transitions due to the SAW create a jump in the magnetization,  
\begin{equation}\label{jump}
\Delta M \propto \int dV F(V) P(V) \propto  f^2P_{\textrm{SAW}}\chi''_{\rm tot}(T,f),
\end{equation}
where $F(V)$ is the volume distribution function satisfying $\int F(V) dV = 1$ and $\chi''_{\rm tot}(T,f) = \int dV F(V) \chi''(V,T,f)$
is the imaginary part of the total susceptibility of the system of NPs. 

One characteristic feature of the spin-rotation mechanism is strong frequency dependence of the effect. It can be traced to the spin-rotation origin of the effective field, ${\bf h} = {\bm \Omega}/\gamma$, with ${\bm \Omega}$ being the time derivative of ${\bm \phi} = \frac{1}{2}{\bm \nabla} \times {\bf u}$.

\begin{figure}[ht]
\begin{center}
\includegraphics[width=9.0cm,angle=0]{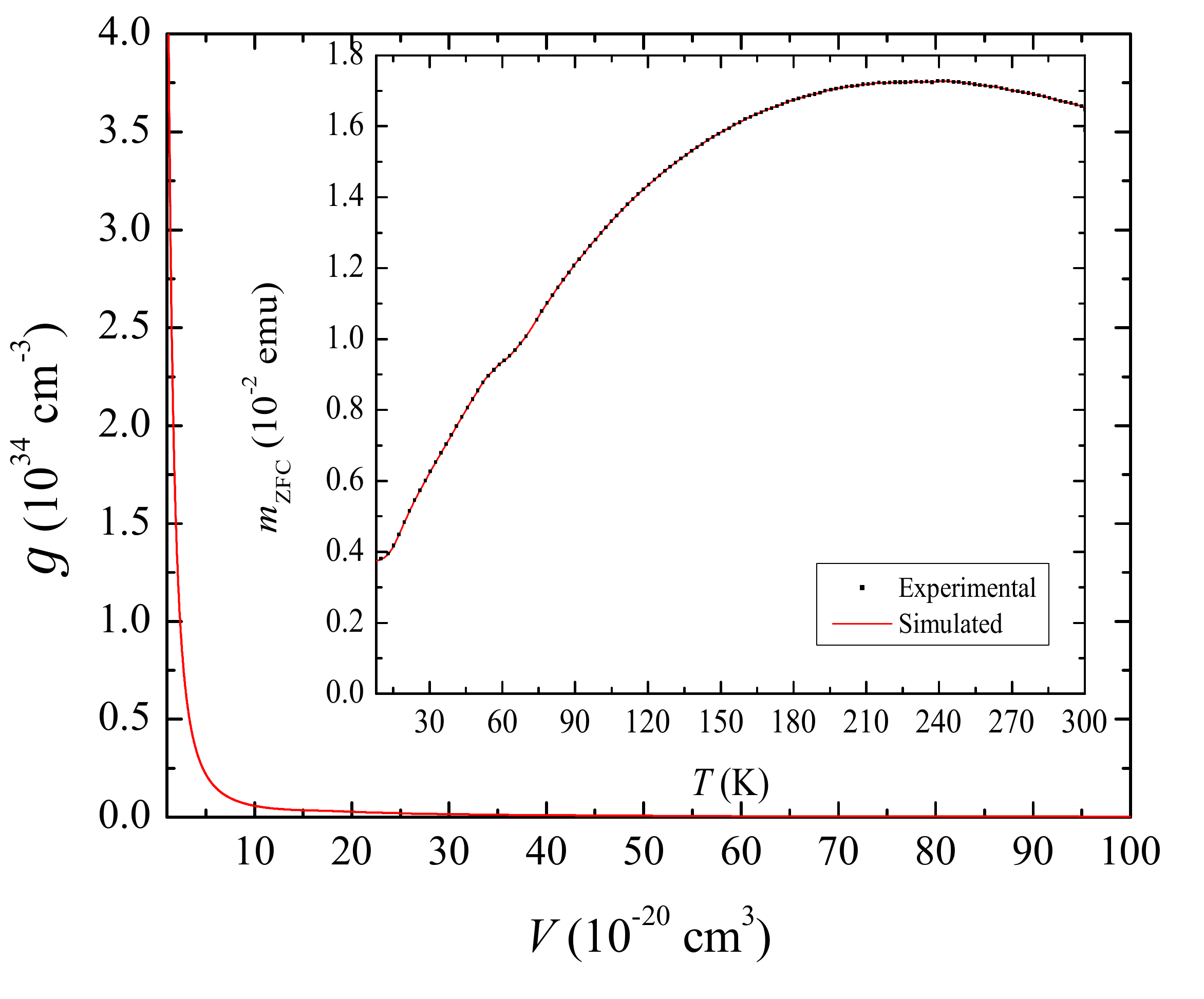}
\caption{Volume distribution of NPs obtained from ZFC measurements. (Inset) ZFC magnetization curve simulated from the plotted volume distribution (red line) and experimental data (black dots), both at $H=100$ Oe.}
\label{vol_distrib}
\end{center}
\end{figure}

\section{Analysis of the data}\label{analysis}

Volume distribution of NPs, $g(V)$, can be obtained from the measured ZFC curve by fitting a Curie-Weiss model for ferromagnetism based on thermal activation of the anisotropy energy barriers,\cite{CT-1998}
\begin{equation}
\label{gV}
g(V)=\frac{2K}{\alpha M_{s}^{2}H}\frac{1}{V^{2}}\frac{\textrm{d}}{\textrm{d}T}\left(Tm_{\textrm{ZFC}}\right)\big[KV/k_{\textrm{B}}\alpha\big],
\end{equation} 
where $\alpha=\ln[t_{\textrm{meas}}/\tau_{0}]$ and $t_{\textrm{meas}}\simeq10$ s represents the measurement time at constant $T$. In what follows we have taken the values $\tau_{0}\sim10^{-9}$ s, $H=100$ Oe and $M_{s}=82$ emu/g, $K=3\cdot10^{5}$ erg/cm$^{3}$ for bulk magnetite. Fig. \ref{Size} implies small number of NPs with diameters bigger than $2R=43$ nm. The cut-off of the volume distribution is even sharper, so that we can safely assume that NPs of volume bigger than $V >V_{\textrm{max}}=\frac{4\pi}{3}R^{3} \sim4.1\cdot10^{-17}$ cm$^3$ do not contribute to the observed effects. On the other hand, the distribution \eqref{gV} diverges asymptotically as $1/V^{1.86}$ in the limit $V\rightarrow0$. Therefore, another cut-off must be imposed on the smallest volume possible, $V_{\textrm{min}}=5.92\cdot10^{-22}$ cm$^{-3}$, which corresponds to the magnetite unit cell.\cite{Comm4}

Fig. \ref{vol_distrib} depicts the volume distribution of NPs in our sample. The inset shows the corresponding simulated ZFC magnetization curve along with the experimental data. Their superimposition determines that both the Curie-Weiss model and our choice of cut-offs capture the essential features of the distribution of magnetite nanoparticles. Integration of $g(V)$ over all volumes yields the estimate $N=9.52\cdot10^{15}$ for the total number of NPs, and therefore the expression $F(V)=g(V)/N$ for the normalized volume distribution.

\begin{figure}[ht]
\begin{center}
\includegraphics[width=9.1cm,angle=0]{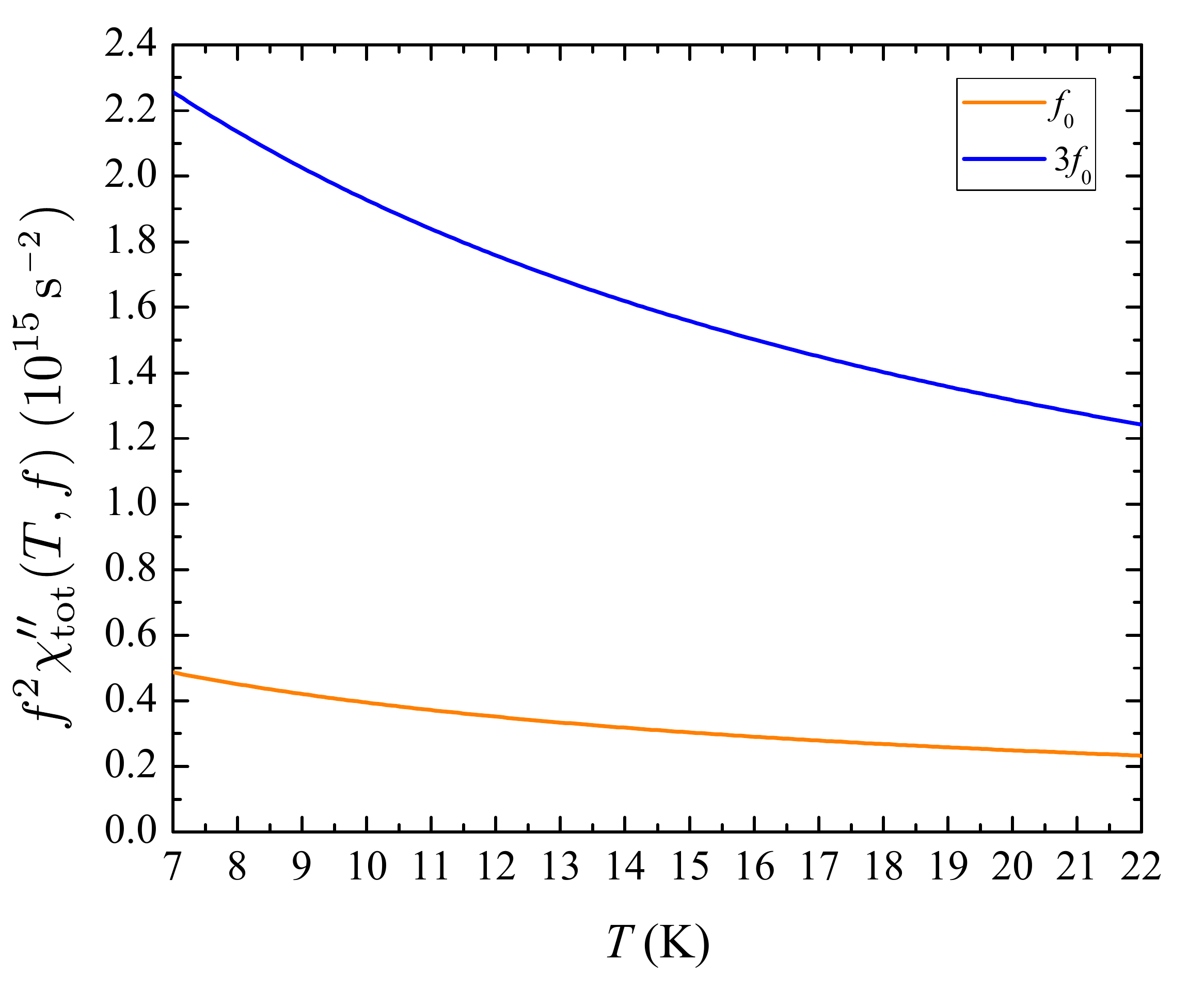}
\caption{Temperature dependence of the prefactor $f^{2}\chi_{\textrm{tot}}''(T,f)$ in Eq. \eqref{jump} at the fundamental frequency of SAW, $f = f_{0}$, (orange line) and the third SAW harmonic, $f = 3f_{0}$ (blue line).}
\label{prefactor}
\end{center}
\end{figure}

\begin{figure}[ht]
\begin{center}
\includegraphics[width=8.6cm,angle=0]{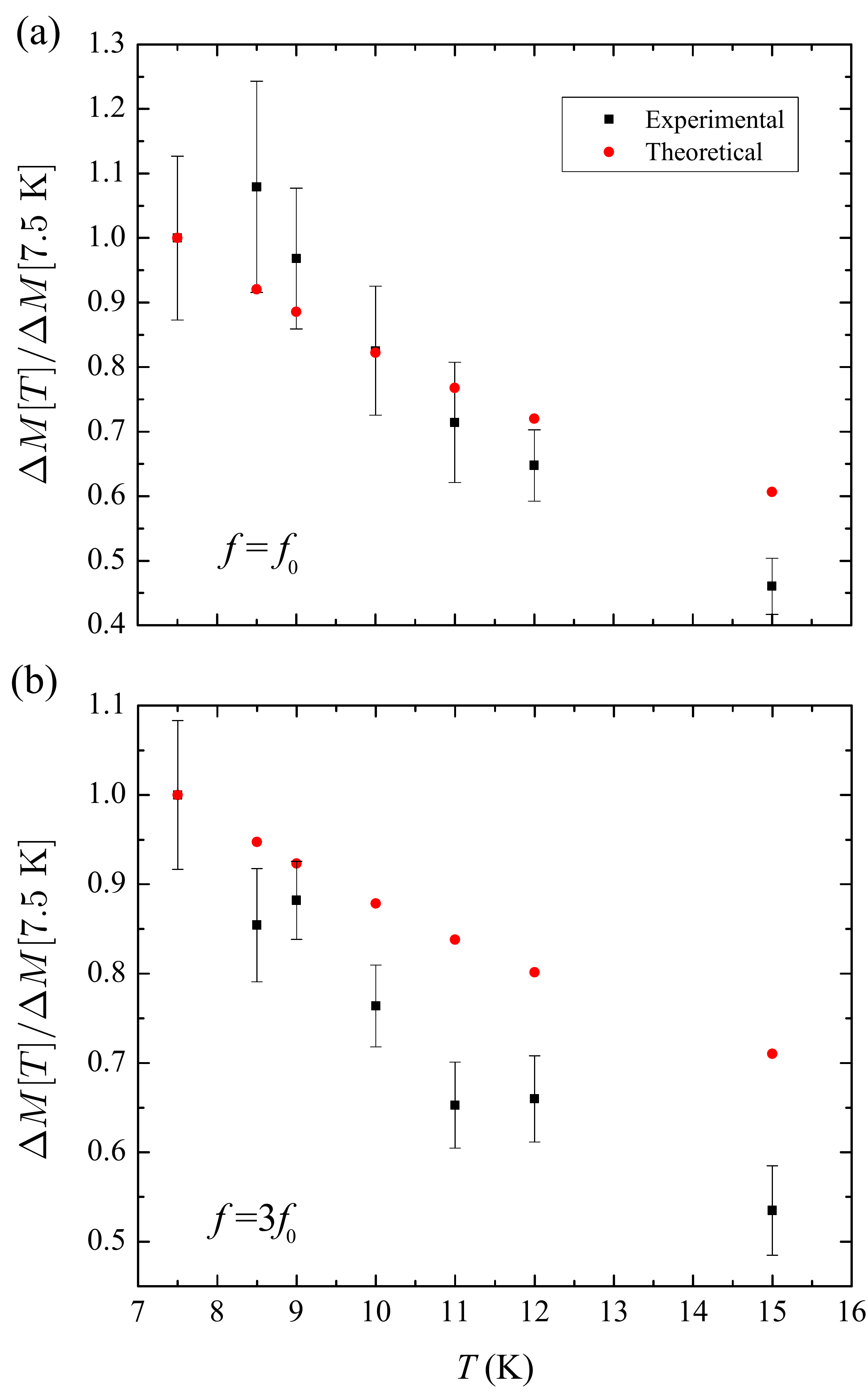}
\caption{Experimental (black) and theoretical (red) thermal dependences of the ratios $\Delta M[T]\big/$ $\Delta M[7.5\textrm{ K}]$ (at a fixed SAW power) at the first [panel (a)] and third [panel (b)] harmonic.}
\label{ratio}
\end{center}
\end{figure}

Calculation of the thermal dependence of the function $f^{2}\chi_{\textrm{tot}}''(T,f)$ is depicted in Fig. \ref{prefactor}. We observe its monotonic decrease with temperature at both fundamental frequency and third harmonic, which agrees well with the trend shown by the slopes of the magnetization jumps---as a function of the SAW power---in Fig. \ref{SAW}. Direct comparison between theory and experiment can be made through the ratios $\Delta M[T]\big/\Delta M[7.5\textrm{ K}]$ at a fixed power, where the unessential prefactors of Eq. \eqref{jump} cancel. Fig. \ref{ratio} shows the experimental and theoretical values of these ratios and, except for a numerical factor, both of them show the same monotonic decreasing dependence.

\section{Discussion}\label{discussion}

We have demonstrated experimentally that surface acoustic waves accelerate transitions between magnetic states of magnetite nanoparticles. Accurate monitoring of the temperature of the sample rules out thermal effects of SAWs that would have accelerated superparamagnetic transitions if the sample were heated. It was previously suggested \cite{Davis-APL2010,Kovalenko-PRL2013,Thevenard-PRB2013,Thevenard-PRB2014,Davis-JAP2015} that magnetostriction due to a SAW can assist magnetization switching in a ferromagnetic layer. It manifests itself in the effect of the tensile stress---compression and/or expansion---on the magnetic anisotropy tensor. If the stress changes the direction of the easy magnetization axis, the magnetic moment relaxes towards the new direction. In a thin layer the tensile stress generated by the SAW acts on the entire layer and it can be strong in materials with large magnetostriction. To achieve strong magnetoelastic coupling between magnetic moments of nanoparticles and the SAW in the piezoelectric substrate, the authors of Ref. \onlinecite{Sampath-Nano2016} grew lithographically flat Co disks on a piezoelectric substrate. In that case the effect of the magnetostriction on the magnetization should be as strong as in a thin Co film deposited on a piezoelectric substrate. 

The Einstein--de Haas mechanism of the magnetization reversal by SAWs suggested in this paper comes from the shear deformation of the substrate and the resulting rotation of the particle as a whole when its size is small compared to the SAW wavelength. It scales linearly with the angular velocity that, in turn, scales as a square of the SAW frequency and linearly with the SAW amplitude. We have shown that the Einstein--de Haas effect is important for SAW frequencies in excess of 100 MHz. At $f=3.5$ MHz used in Ref. \onlinecite{Sampath-Nano2016} it can be safely neglected. While it is possible that magnetostriction also contributes to the effect observed in our experiment, this contribution is hard to estimate for magnetite particles with weak magnetoelastic coupling, touching the substrate at random points. It is obvious, however, that it must be small compared to the effect in flat magnetic islands of materials with high magnetoelastic constants, such as Co, grown on a piezoelectric substrate with the purpose to maximize the effect of magnetostriction.\cite{Sampath-Nano2016} On the contrary, the novel mechanism suggested by us, that contributes dominantly at high frequencies, should definitely accomplish the purpose. 

The beauty of the Einstein--de Haas mechanism is that the estimate of its strength does not require any knowledge of the parameters of the particles. Instead, it is determined by the angular velocity of the rotation of the substrate---due to the SAW---at the position of the nanoparticle, which depends on the parameters of the SAW but not on the parameters of the nanoparticle. When the frequency of the SAW is above 100 MHz and its amplitude is in the ballpark of 0.01\% of the wavelength---which has been realized in our experiment---, the effective ac magnetic field in the coordinate frame of the particle exceeds 0.1 Oe. It is well known that such ac fields are sufficient to reverse magnetic moments through consecutive absorption of photons of spin 1, see Ref. \onlinecite{CGC-PRB2013} of the manuscript and references therein. In our case the surface phonons get consecutively absorbed by the magnetic moment, not the photons. Notice that this mechanism has nothing to do with the reversal of the magnetic moment by the dc magnetic field that exceeds the magnetic anisotropy field; the required amplitude of the ac field is very small compared to the anisotropy field. 

By relating the effect of SAWs to the effect of an ac magnetic field we have been able to reproduce the experimental dependences on the power and the frequency of the SAW, as well as on temperature. Notice that there are no existing models based upon magnetostrictive effects capable of reproducing our experimental results. Our observation opens the way of manipulating magnetic moments of nanoparticles by SAWs. The advantage of this method, vs manipulating magnetic moments by the ac magnetic field, is the five orders of magnitude smallness of the wavelength of the SAW as compared to the wavelength of the electromagnetic radiation of the same frequency. This could, in principle, open the way of individual manipulation of densely packed nanoparticles by SAWs.

\section{Acknowledgements}
Preparation of NPs and their characterization by TEM and XRD methods were performed by das-Nano, S.L, enterprise in Pamplona, Spain. The work of E.M.C. has been supported by the PSC-CUNY Grant No. 69057-0047. J.T. and N.S. acknowledge funding from MINECO through MAT2015-69144. R.Z. thanks Fundaci\'{o}n Ram\'{o}n Areces for support through a postdoctoral fellowship within the XXVII Convocatoria de Becas para Ampliaci\'{o}n de Estudios en el Extranjero en Ciencias de la Vida y de la Materia. J.C.-de la R. acknowledges Ajuts a la Doc\`{e}ncia i a la Recerca (ADR) given by Universitat de Barcelona, and the research group DIOPMA (2014 SGR 1543). N.S. acknowledges funding from SURDEC through the research training grant FI-DGR. P.V.S. and A.H.-M. thank W. Seidel and S. Rauwerdink for assistance in the preparation of the acoustic delay lines.

\end{document}